\documentclass[12pt,fleqn]{article}
\usepackage{a4,epsfig,epic}

\newcommand{\sibe}{\ifmmode\sigma_{\beta}
                   \else$\sigma_{\beta}$\fi}
\newcommand{\sicha}{\ifmmode\sigma_{\theta}
                   \else$\sigma_{\theta}$\fi}

\newcommand{\AmS}{{\protect\the\textfont2
   A\kern-.1667em\lower.5ex\hbox{M}\kern-.125emS}}

\begin{document}

\begin{titlepage}

\large
\centerline {\bf \Large The Concurrent Track Evolution 
Algorithm: }
\vskip 3mm
\centerline {\bf \Large Extension for Track Finding in the Inhomogeneous}
\vskip 3mm
\centerline {\bf \Large Magnetic Field of the HERA-B Spectrometer}
\normalsize
 
\vskip 2.0cm
\centerline {Rainer Mankel~\footnote{Email: {\tt mankel@ifh.de}}}
\centerline {\it Institut f\"ur Physik, Humboldt Universit\"at zu
  Berlin}
\centerline{\it Invalidenstr 110, D--10115 Berlin, Germany}
\vskip 1.0cm
\centerline{and}
\vskip 1.0cm
\centerline {Alexander Spiridonov~\footnote{
                        Email: {\tt spiridon@ifh.de}.} 
                        \footnote{Permanent address: 
                          Institute for High Energy 
                            Physics, 142284 Protvino, Russia. 
}}
\centerline {\it DESY Zeuthen}
\centerline {\it Platanenallee 6, D--15738 Zeuthen, Germany}

\vskip 2.0cm

 
\centerline {\bf Abstract}
\vskip 0.8cm

  The Concurrent Track Evolution method, which was introduced in a
  previous paper~\cite{conc}, has been further explored by applying it
  to the propagation of track candidates into an inhomogeneous
  magnetic field volume equipped with tracking detectors, as is
  typical for forward B spectrometers like HERA-B or LHCb.  Compared
  to the field-free case, the method was extended to three-dimensional
  propagation, with special measures necessary to achieve fast
  transport in the presence of a fast-varying magnetic field.  The
  performance of the method is tested on HERA-B Monte Carlo events
  with full detector simulation and a realistic spectrometer geometry.
\vskip 1.0cm

\vfill
\end{titlepage}

\newpage

\section{Introduction}
Dedicated B physics experiments for hadronic interactions, as
HERA-B~\cite{herabproposal} and LHCb~\cite{lhcb} are designed as
forward spectrometers because of the kinematics of the $b$ hadrons,
whose decay particles are produced with a large Lorentz boost.  Even
with highly granular tracking devices, as honeycomb drift chambers and
strip detectors, the large track density leads to high cell
occupancies, which can range up to 20\% in the ``hottest'' parts of
the HERA-B outer tracker. Under these circumstances, pattern
recognition becomes a crucial issue.

\begin{figure}[ht]
\begin{center}
\unitlength1.5cm
\begin{picture}(9,8)

\dashline{0.1}(0.6,4.20)(8.8,4.20)
\put(8.8,4.20){\vector(1,0){0.1}}
\put(9.0,4.00){\makebox(0,0)[l]{z}}

\dashline{0.1}(0.6,4.20)(0.6,5.)
\put(0.6,5.35){\vector(0,1){0.1}}
\put(0.4,5.45){\makebox(0,0)[b]{x}}

\put(1.,1.7){\epsfig{file=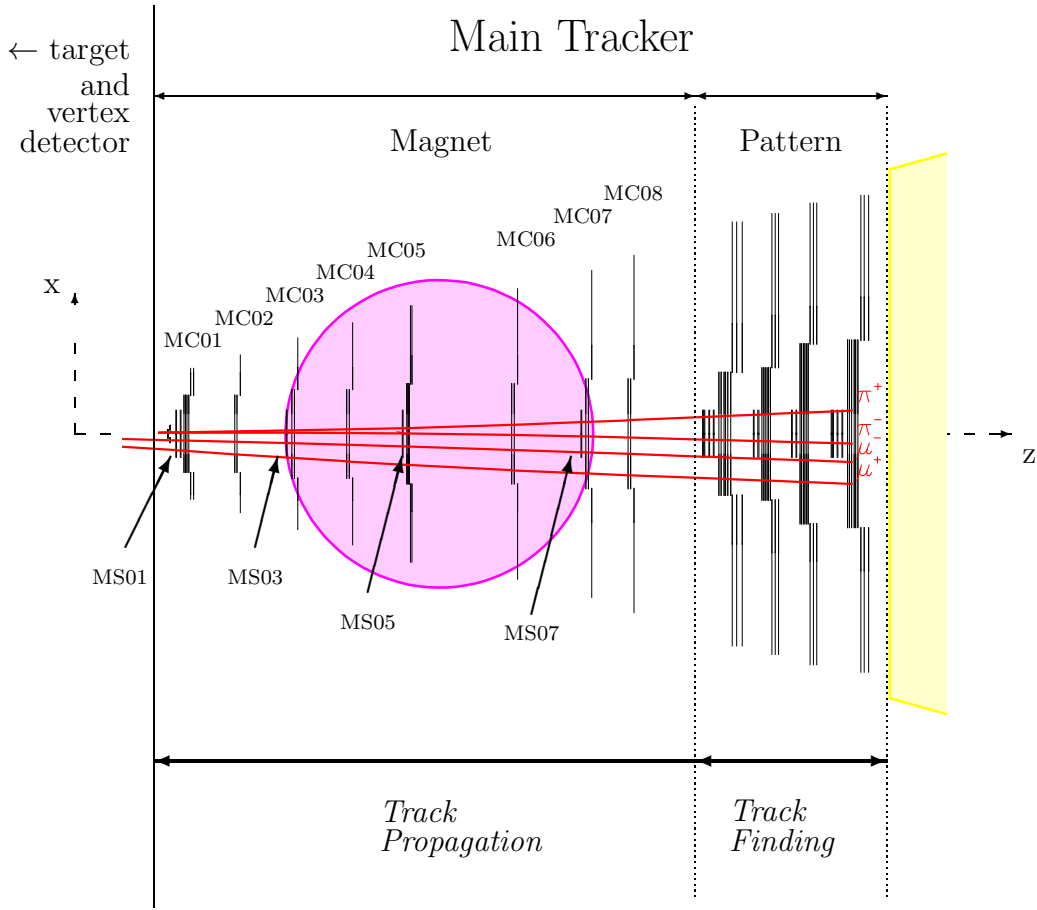,width=11cm}}

\put(1.3,0){\line(0,1){8}}
\dottedline{0.05}(6.10,0.1)(6.10,7.1)
\dottedline{0.05}(7.80,0.1)(7.80,7.1)

\put(1.30,7.2){\vector( 1,0){4.80}}
\put(6.10,7.2){\vector(-1,0){4.80}}
\put(6.10,7.2){\vector( 1,0){1.70}}
\put(7.80,7.2){\vector(-1,0){1.70}}

\put(0   ,6.7){\shortstack[r]{$\leftarrow$ target \\ and 
        \\vertex\\ detector}}
\put(5   ,7.6){\makebox(0,0)[b]{\Large Main Tracker}}

\put(3.85, 6.9){\makebox(0,0)[t]{Magnet}}
\put(6.95, 6.9){\makebox(0,0)[t]{Pattern}}

\thicklines
\put(6.10,1.3){\vector(-1,0){4.8}}
\put(6.10,1.3){\vector(1,0){1.7}}
\put(7.80,1.3){\vector(-1,0){1.7}}

\put(3.3 ,0.5){\shortstack[l]{\it Track\\ \it Propagation}}
\put(6.4,0.5){\shortstack[l]{\it Track\\ \it Finding}}

\put(1.65, 5.1){\makebox(0,0)[t]{\scriptsize MC01}}
\put(2.10, 5.3){\makebox(0,0)[t]{\scriptsize MC02}}
\put(2.55, 5.5){\makebox(0,0)[t]{\scriptsize MC03}}
\put(3.00, 5.7){\makebox(0,0)[t]{\scriptsize MC04}}
\put(3.45, 5.9){\makebox(0,0)[t]{\scriptsize MC05}}
\put(4.60, 6.0){\makebox(0,0)[t]{\scriptsize MC06}}
\put(5.10, 6.2){\makebox(0,0)[t]{\scriptsize MC07}}
\put(5.55, 6.4){\makebox(0,0)[t]{\scriptsize MC08}}

\put(1.05, 3.2){\vector(1,2){0.40}}
\put(1.00, 3.0){\makebox(0,0)[t]{\scriptsize MS01}}

\put(2.20, 3.2){\vector(1,4){0.2}}
\put(2.20, 3.0){\makebox(0,0)[t]{\scriptsize MS03}}
\put(3.20, 2.8){\vector(1,4){0.30}}
\put(3.20, 2.6){\makebox(0,0)[t]{\scriptsize MS05}}
\put(4.65, 2.6){\vector(1,4){0.35}}
\put(4.65, 2.5){\makebox(0,0)[t]{\scriptsize MS07}}

\end{picture}

\end{center}

\caption{
Layout of the HERA-B main tracking system (top view) in front of the
RICH. The proton beam enters from the left side. The tracking areas
used for track finding and track propagation (see text) are indicated
at the bottom. The shaded areas are the pole shoes of the magnet
(left) and the first part of the RICH (right). This paper concentrates
on the track propagation from the pattern tracker into the {\it magnet
tracker}.}
\label{trackerlayout-mag}
\end{figure}

In a previous paper~\cite{conc}, the {\it Concurrent Track Evolution}
strategy has been presented as a method of track finding in a {\it
multi-planar} tracker geometry. Concurrent Track Evolution is a {\it
track following method} based on the Kalman
filter~\cite{kalman,fruehwirth,billoir,kalmannote}, which evaluates
the {\it available paths} for a track candidate {\it concurrently} to
find the optimal solution, but also keeps the combinatorics at a
reasonable level (see~\cite{conc} for details). The method has been
successfully applied to find initial track segments in the geometry of
the HERA-B {\it pattern tracker}, which consists of four superlayers
of inner and outer tracking devices in the field-free region between
the magnet and the RICH. The layout of the HERA-B spectrometer is
shown in fig.~\ref{trackerlayout-mag}. A more detailed explanation of
the HERA-B tracker and the reconstruction strategy can be found
in~\cite{conc}.

\begin{figure}[hbp]
\begin{center}
\unitlength1cm
\begin{picture}(10,19)
\put(0.0,18.0){\makebox(0,0)[t]{\huge (a)}}
\put(1.0,9.5){\epsfig{file=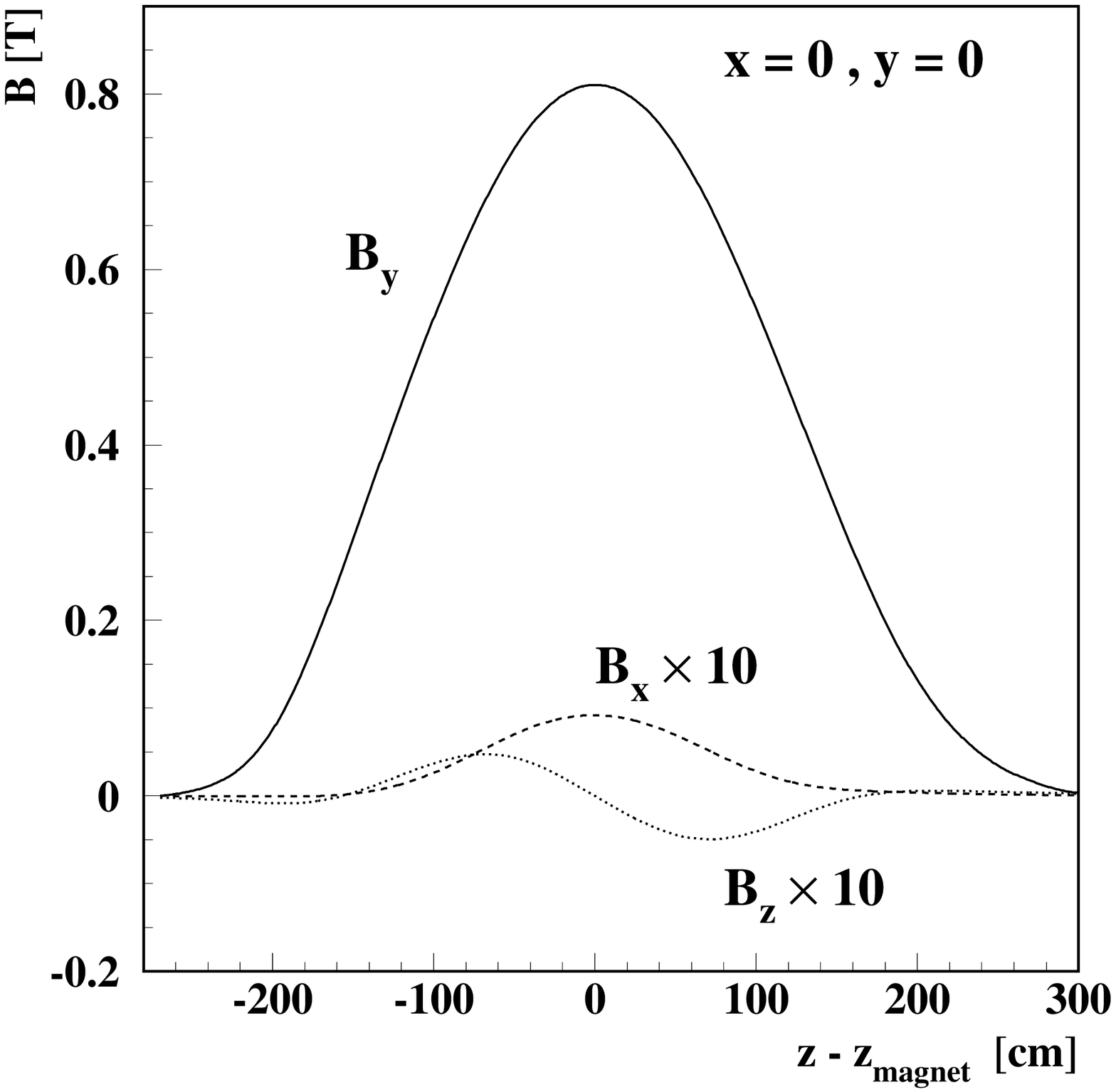,width=9cm}}
\put(0.0,8.5){\makebox(0,0)[t]{\huge (b)}}
\put(1.0,0.0){\epsfig{file=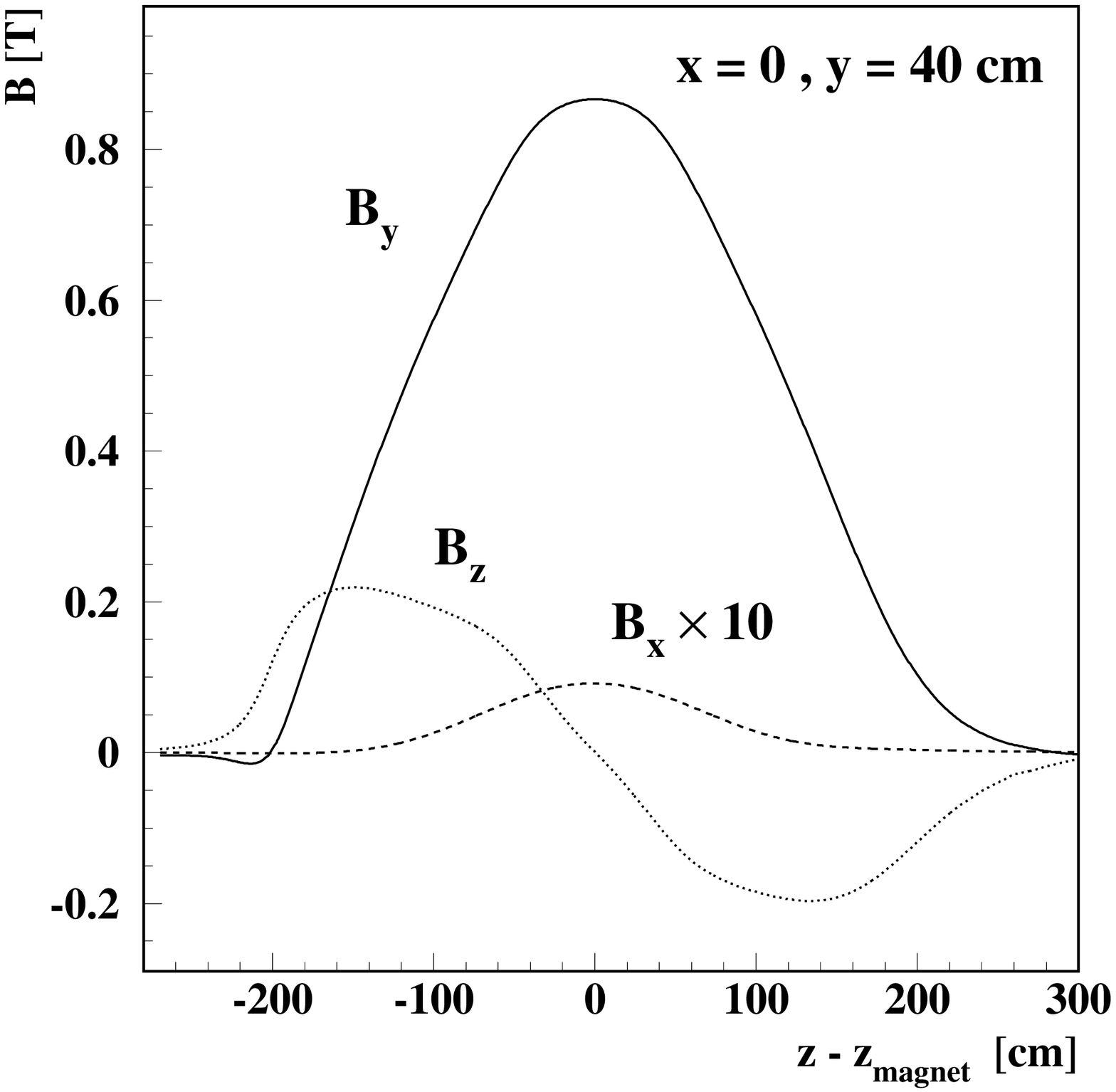,width=9cm}}
\end{picture}
\caption{Magnetic field components as function of the $z$ coordinate in 
the magnet center (a) and at a vertical displacement of 40~cm (b). The
$z$ coordinate is given relative to the magnet center at
$z_{magnet}=450$~cm.}
\label{field}
\end{center}
\end{figure}

The next step in the HERA-B track reconstruction is the track
propagation through the magnetic field. Typical sizes of the magnetic
field components $B_x$, $B_y$ and $B_z$ as a function of $z$ are
displayed in fig.~\ref{field}. The main bending component ($B_y$) has
a bell-shaped dependence on $z$, and there is clearly no significant
subset of the magnet tracking system in which the field can be
regarded as homogeneous. Moreover, the $x$ and $z$ components are
sizeable, making it impossible to restrict pattern recognition to
projections while making full use of the track model. The necessity to
deal with five track parameters simultaneously, and the additional
hits generated by spiralling particles make standalone track finding
within the magnet a problematic task. Instead, the HERA-B track
reconstruction concept operates by first finding straight-line track
segments in the pattern tracker (described in~\cite{conc}) and then
{\it propagating} them upstream through the magnet area. In a
subsequent step, these track candidates will have to be matched to the
track segments in the silicon vertex detector. A considerable fraction
of the $K^0_S$ from the {\it golden decay} mode $B^0 \rightarrow
J/\psi K^0_S$ decays too late to produce a sufficient number of hits
in the acceptance of the vertex detector, so that the vertex must be
reconstructed from the main tracker information.

Compared to pattern recognition {\it from scratch} in the magnet, this
{\it propagation} approach has the advantage that only one track
parameter, the momentum, is poorly known and must be fitted during the
propagation. There are however several challenges involved:
\begin{itemize}
\item Due to the non-linear nature of the fit, iteration, in general, 
may be required to obtain the optimal result. In particular, it might
happen that upstream hits are not found within the predicted error
margin because a wrong initial momentum value in the downstream
section caused derivatives to be taken at the wrong positions in
parameter space. Iteration of the various track candidates in the
context of Concurrent Track Evolution, on the other hand, would
multiply the computational effort to a worrisome degree.
\item The inhomogeneity of the field necessitates a much more costly,
numerical transport of the track parameters than in the field-free
case. The Kalman filter requires also the transport of the covariance
matrices, for which the derivative matrices of the transported versus
the original parameters must be calculated.
\item The different superlayer structures of outer and inner magnet
tracker represent additional challenges to the {\it navigation}
concept.
\end{itemize}

\section{Geometry}
As an example, we will use the current HERA-B detector layout as it is
presently implemented in the geometry definition of the
experiment. This layout is shown in fig.~\ref{trackerlayout-mag}. On
the right-hand side, four superlayers of inner and outer tracking
devices form the {\it pattern tracker}, which is used for finding the
initial straight-line track segments. The left part shows the {\it
magnet tracker} through which these segments have to be
propagated. The magnet tracker consists of eight superlayers of outer
({\tt MC01...MC08}) and four superlayers of inner tracker ({\tt MS01,
MS03, MS05, MS07}). The outer tracker is based on honeycomb drift
chamber technology with $5$ and 10~mm diameters, with an assumed
resolution of $200\mu $m (see
\cite{conc,herabdesign} for further details). Each outer tracker
superlayer in the magnet consists of three tracking layers with the
orientations $+\alpha$, 0$^\circ$, $-\alpha$ with a stereo angle of
$\alpha \approx 100$~mrad, except for {\tt MC01} which has four layers
in a 0$^\circ$, $-\alpha$, 0$^\circ$, $+\alpha$ arrangement to satisfy
requirements of second level triggering, and {\tt MC05} which consists
only of a single 0$^\circ$ layer because of space restrictions inside
the magnet yoke. The inner tracker is built of micro-strip gaseous
chambers (MSGC) with a typical resolution of $80\mu $m. Each
superlayer consists of one zero degree and one stereo layer with a
stereo angle of $+100$ or $-100$~mrad, alternatingly, except for {\tt
MS01} which has four layers with the sequence 0$^\circ$,
$-\alpha$, $+\alpha$, 0$^\circ$. The sector structure of the magnet
tracker superlayers is similar to that of the pattern tracker
(see~\cite{conc}).

Compared to the setup shown in~\cite{conc}, the stations {\tt MC09}
and {\tt MS09} have been removed because they are already in an area
of weak magnetic field and are not expected to improve the track
parameters already measured by the pattern tracker
significantly\footnote{It is also foreseen to remove {\tt MC07}}. The
gaps in the second half of the magnet tracker are reserved for pad
chambers designed to generate a pretrigger for high momentum tracks,
which is intended to select $B^0 \rightarrow \pi^+\pi^-$ decays.  In
front of {\tt MS01}, an additional silicon strip detector superlayer
is foreseen to cover the innermost angular region. Both the pad
chambers and the silicon superlayer were not used in this study.

\section{Method}

\subsection{Seed}
\begin{figure}
\begin{center}
\unitlength1cm
\begin{picture}(15,15)
\put(0.0,14.1){\makebox(0,0)[t]{\large (a)}}
\put(0.5, 7.5){\epsfig{file=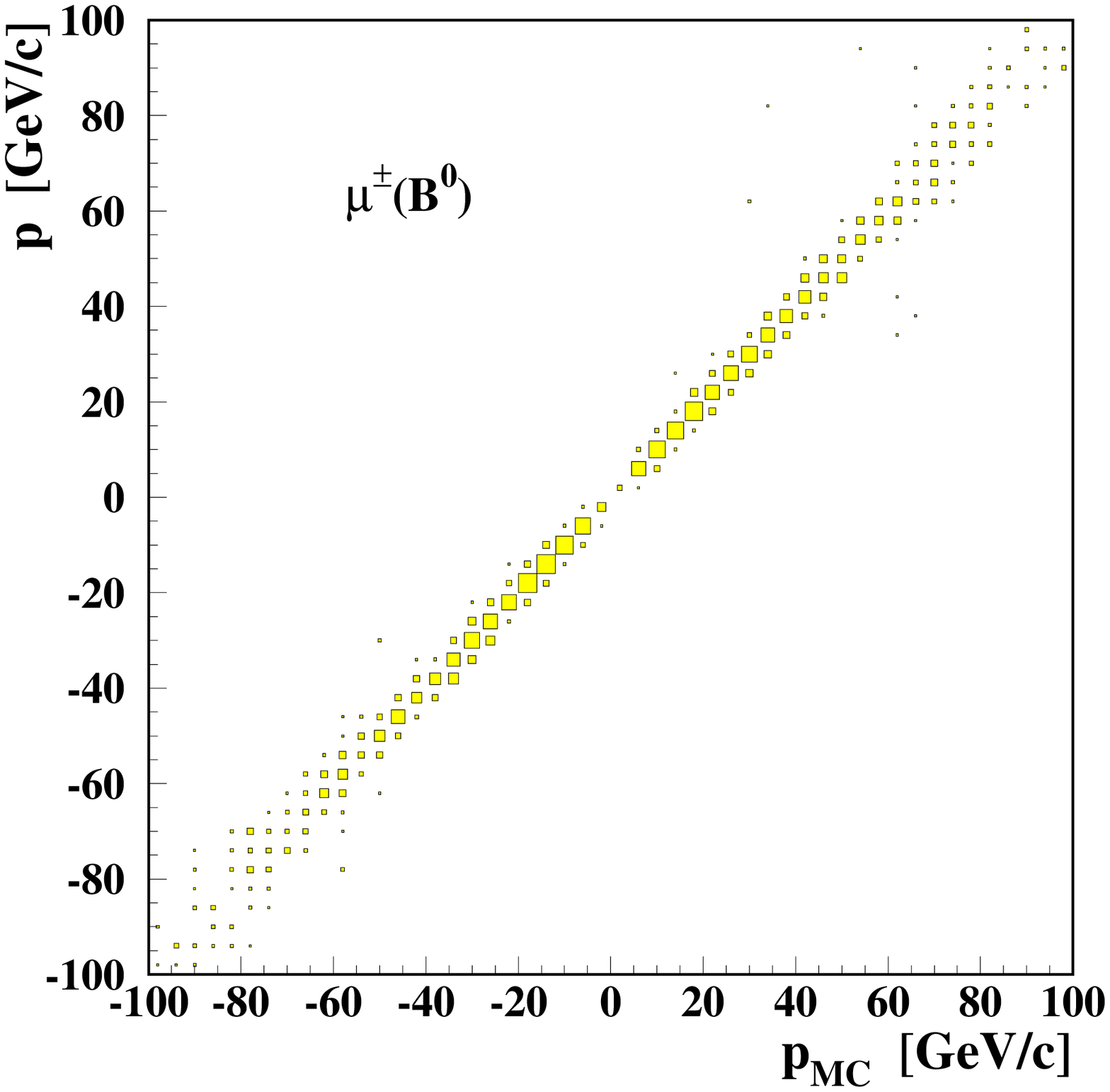,width=7cm}}
\put(7.5,14.1){\makebox(0,0)[t]{\large (b)}}
\put(8.0, 7.5){\epsfig{file=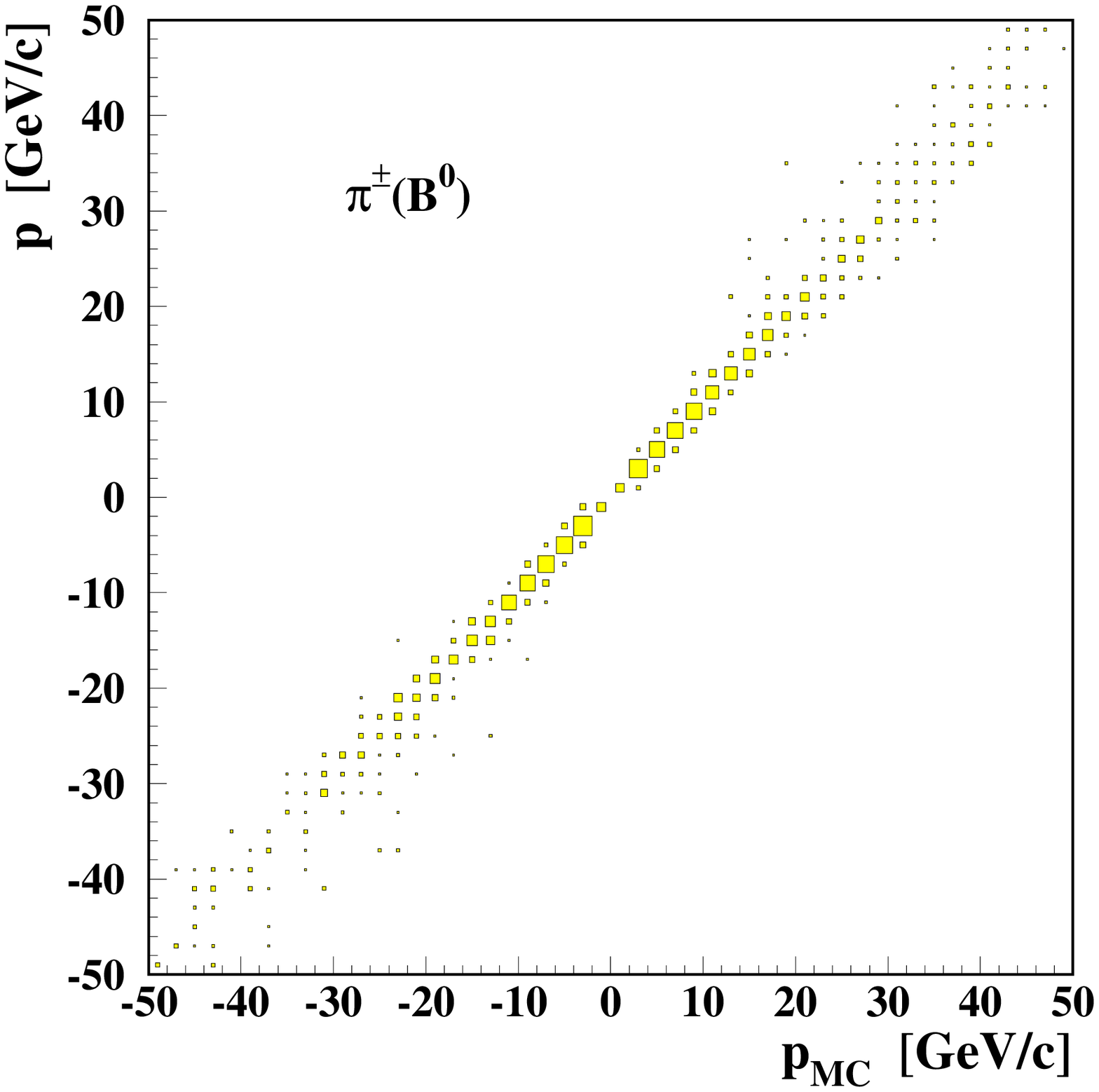,width=7cm}}
\put(0.0, 6.5){\makebox(0,0)[t]{\large (c)}}
\put(0.5, 0.0){\epsfig{file=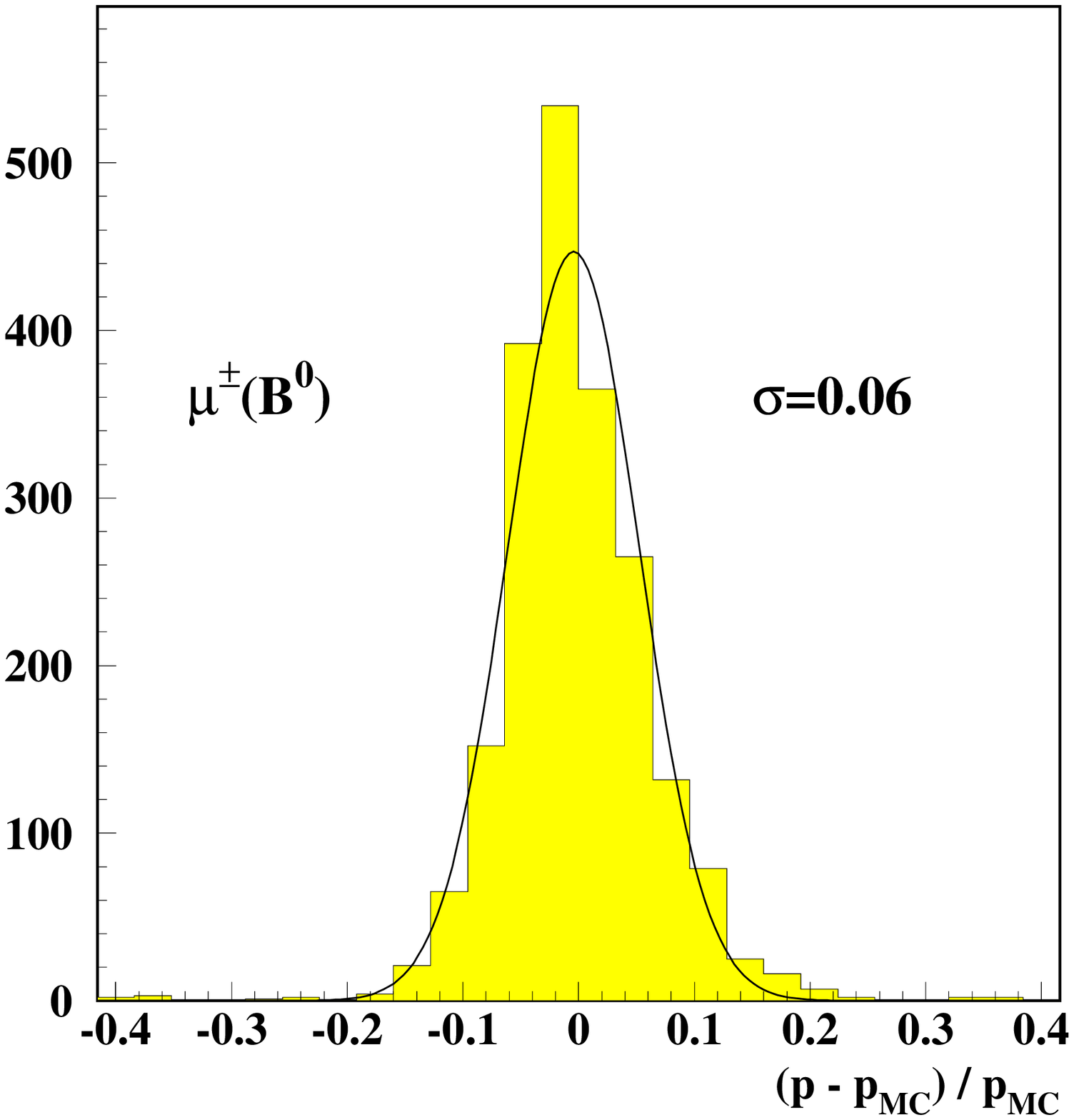,width=7cm}}
\put(7.5, 6.5){\makebox(0,0)[t]{\large (d)}}
\put(8.0, 0.0){\epsfig{file=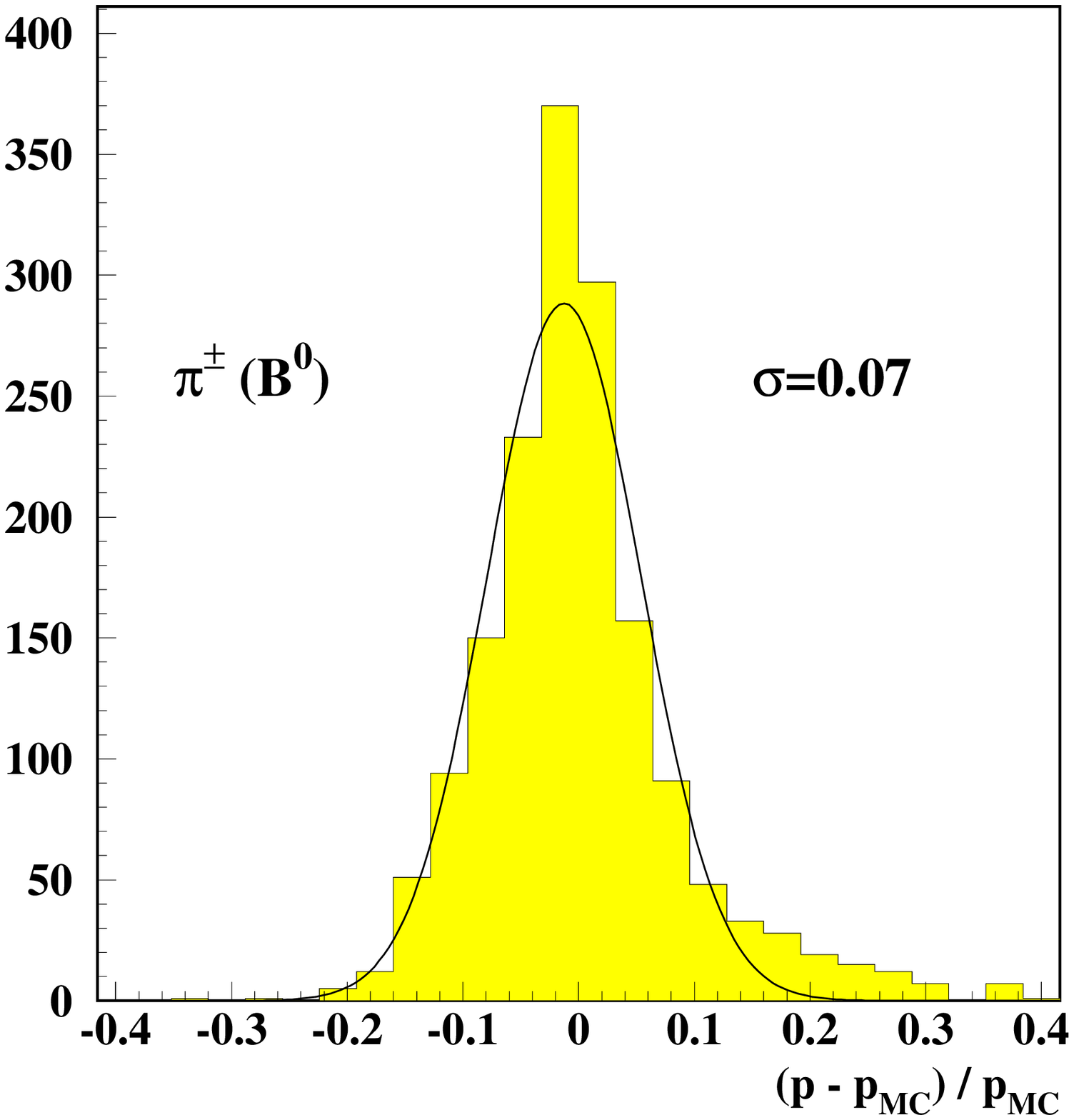,width=7cm}}
\end{picture}
\end{center}
\caption{Correlation (a--b) between the momentum estimate (signed
according to charge) from the apparent track deflection (see text),
and the corresponding Monte Carlo value for muons (a) and pions (b)
from the golden B decay.  (c--d) show the distribution of the relative
momentum error for the muons (c) and pions (d).}
\label{pEstimate}
\end{figure}

A track candidate found in the pattern tracker provides the starting
seed, which gives already rather precise estimates of the parameters
$x$, $y$, $t_x = p_x/p_z$ and $t_y = p_y/p_z$ at the entrance of the
pattern tracker, which is at $z\approx 700$~cm. We use the inverse
momentum signed according to charge, $Q/p$, as momentum parameter
because it is most convenient to use in the numerical treatment of
transport within the magnetic field. Since iterations are to be
avoided, it is essential to have already an estimate of the momentum
parameter which is good enough to give reliable parameter
derivatives. Such an estimate can be obtained from the visible
deflection by the magnetic field assuming that the track comes from
the target region:
\begin{displaymath}
        \frac{Q}{p} =
        \frac{\left[ x -  (z - z_{target}) t_x \right]\ 
               \sqrt{1 + t_x^2}}
          {\int B_yd\ell \cdot  (z_{magnet} - z_{target})
               \sqrt{1 + t_x^2 + t_y^2}}
\end{displaymath}
where $\int B_yd\ell $ is the average field integral of the main
bending component, $z_{magnet}$ and $z_{target}$ are the $z$
coordinates of magnet center and target and $x$ and $t_x$ are the
impact parameter and track angle in the bending plane as given at $z$
by the seed. The quality of this estimate for muons and pions from the
golden B decay is shown in fig.~\ref{pEstimate}. The estimate gives
obviously reliable results in most cases. The relative precision is
$\approx 6\%$ in the muon and $\approx 7\%$ in the pion case, in the
latter the estimate is slightly diluted by the displacement of the
$K^0_S$ decay vertex which effects the tail in
fig.~\ref{pEstimate}d. The initial covariance matrix diagonal element
of the momentum parameter is set accordingly.

\subsection{Modifications to the Concurrent Track Evolution method}
After the initial seed definition from the pattern tracker, the
Concurrent Track Evolution mechanism~\cite{conc} propagates the
candidate as far as possible upstream within the magnet tracker,
employing the Kalman filter technique. This is achieved by a
succession of {\it growth cycles}, which explore the possible
continuations from layer to layer. A certain number of branches is
propagated concurrently to explore the available paths. We summarize
here briefly the steps which are explained in detail in~\cite{conc}:
\begin{enumerate}
\item Based on the position of the last hit of a candidate, generate a 
list of detector plane parts in which the next logical hit is
expected, using the {\it domain navigation method} described
in~\cite{conc}. Define a search window in each plane, using the Kalman
filter prediction from the last hit.

\item Create a new branch for each hit found in each search window.
\item Filter each hit and accept the new branch if the
$\chi^2$ contribution is not too large.
\item At the end of each growth cycle, locate the {\it best} candidate
according to a quality estimator function. Discard all branches whose
quality estimates differ by more than a given maximum value, keep only
the ${\cal R}_{max}$ best candidates.
\item Repeat the steps above, until no further growth is
possible. This may be either because the end of the tracking system is 
reached, or because no further suitable hits are found.
\item Select the best remaining candidate, and store it if its hit 
count exceeds a certain minimum.
\end{enumerate}

Compared to the algorithm decribed in~\cite{conc},
the following changes and extensions have been made:
\begin{enumerate}
\item The algorithm is applied in three dimensions. The same navigation
strategy as in~\cite{conc} is used, but a domain window is
immediately discarded if it is not intersected by the extrapolated
track candidate.
\item Hits are considered for filtering if they are within either
three standard deviations or 2~cm of the predicted projected
coordinate $u = x \ \cos \alpha_{st} - y\ \sin \alpha_{st}$ (where
$\alpha_{st}$ is the stereo angle).
\item After filtering, the new candidate is accepted if the $\chi^2$
contribution is less than 16.
\item In the quality estimator $Q$ (see~\cite{conc}), which is calculated
from the number of hits and their $\chi^2$ contributions, the weight
of the latter ($w_{\chi^2}$) is set to 0.05.
\item If at least one new track candidate is obtained through
successful continuation, its parent track candidate is excluded from
further propagation, This is unlike the strategy in the
projection-based tracking in~\cite{conc}: without knowledge of the
orthogonal coordinate, the probability to pick up a wrong hit within
the prediction window was higher, so that the possibility of a missing
hit ({\it fault}) had to be considered even if a continuation hit had
been found.
\item Since the magnet part of the inner tracker has only about half
as many layers as the outer tracker, the inner tracker hits are
weighted by a factor of two when the number of hits is calculated. A
track is accepted for storage if its weighted number of hits, acquired
in the magnet tracker is at least 5.
\end{enumerate}
The maximum number of missing hits in sequence ($N^{max}_{Faults}$) is
kept at 2, and the maximum number of track candidates treated
concurrently (${\cal R}_{max}$) remains at 5.

\subsection{Transport of track parameters within the magnetic field}
Compared to the field-free case discussed in~\cite{conc}, where
two-parameter states are transported linearly, the transport within
the inhomogeneous field is much more critical because there are five
track parameters involved, and both the parameters and their covariance 
matrix must be transported. Regarding the number of track
candidates which appear in the course of Concurrent Track Evolution,
very efficent numerical procedures are needed in order not to exceed
practical cpu time limitations.

It has already been shown that the precision requirements of HERA-B
tracking can be met using a fifth-order Runge-Kutta method with
adaptive step size control~\cite{rk}, a procedure which has also been
successfully used for track fitting~\cite{rangerfit}. For the track
propagation problem discussed in this paper, we have applied the
following flexible strategy (described in detail in~\cite{tracing}) to
keep the computational effort at a minimum:
\begin{description}
\item[short distances:] for $\delta z <$20~cm, a parabolic expansion of 
the trajectory, based on the field vector at the starting point, has been
used.
\item[medium distances:] for 20~cm$<\delta z <$60~cm, a classical
fourth-order Runge-Kutta method~\cite{recipes} has been applied.
\item[long distances:] for $\delta z >$60~cm, a fifth-order
Runge-Kutta method with adaptive step size control~\cite{recipes} has
been employed.
\end{description}

Since the precision for the derivatives of the transported parameters
($\tilde x$, $\tilde y$, $\tilde t_x$, $\tilde t_y$, $Q/\tilde p$) with
respect to the untransported ones ($x$, $y$, $t_x$, $t_y$, $Q/p$) is less
critical than the precision of the parameters themselves, the sizes of the
derivatives were studied and their influence on the track finding
result was carefully investigated~\cite{tracing}. It was found that
the derivatives with respect to the first four parameters can be
approximated by
\begin{displaymath}
  \partial (\tilde x,\tilde y,\tilde t_x,\tilde t_y,Q/\tilde
          p) / \partial x = (1, 0, 0, 0, 0)
\end{displaymath}
\begin{displaymath}
  \partial (\tilde x,\tilde y,\tilde t_x,\tilde t_y,Q/\tilde
          p) / \partial y =  (0, 1, 0, 0, 0)
\end{displaymath}
\begin{displaymath}
  \partial (\tilde x,\tilde y,\tilde t_x,\tilde t_y,Q/\tilde
          p) / \partial t_x = (\tilde z - z, 0, 1, 0, 0)
\end{displaymath}
\begin{displaymath}
   \partial (\tilde x,\tilde y,\tilde t_x,\tilde t_y,Q/\tilde
          p) / \partial t_y = (0, \tilde z - z, 0, 1, 0)
\end{displaymath}
without loss in efficiency but with large gain in speed. The
derivatives with respect to the momentum parameter are approximated as
\begin{displaymath}
   \partial (\tilde x,\tilde y,\tilde t_x,\tilde t_y,Q/\tilde
          p) / \partial (Q/p) = (\partial \tilde x/\partial (Q/p), 0,
                              \partial \tilde t_x/\partial (Q/p), 0, 1)
\end{displaymath}
where $\partial \tilde x/\partial (Q/p)$ and 
$\partial \tilde t_x/\partial (Q/p)$
are defined by differential equations which are solved together with
the integration of the equations of motion by the Runge-Kutta method.

These approximations lead to a derivative matrix whose elements
contain mainly ones and zeroes. This property is used to reduce
strongly the number of computations for the transport of the
covariance matrix.

The propagation of electrons and positrons within the magnetic field
is complicated by the energy loss through bremsstrahlung. An
additional radiative energy loss correction according to the method of
Stampfer et al.~\cite{stampfer,kalmannote} has been applied, which was
found to improve the magnet tracking efficiency for electrons from the
golden $B$ decay by 0.5\% with respect to the number of reference
tracks (sec.~\ref{sec:perfEst}).

\section{Performance}

\subsection{Event sample}
Samples of 1219 interactions of the type 
$pAl \rightarrow B^0+X$, $B^0 \rightarrow J/\psi K^0_S
\rightarrow (\ell^+ \ell^-) (\pi^+ \pi^-)$, with 915 decays into
the muon and 304 into the electron channel, and 2000 unbiased
inelastic $pAl$ interactions were generated~\cite{lund} and passed
through the full HERA-B detector simulation~\cite{hbgean} to
investigate the performance. Both the event generation and the
detector digitization were done in the same manner as described
in~\cite{conc}.

\subsection{Performance estimators}
\label{sec:perfEst}
As in~\cite{conc}, efficiencies are calculated for {\it reference
tracks} which are a track class defined by both the geometrical
acceptance and the physics interest of the experiment. Since a track
segment from the pattern tracker is needed as a starting point, a {\it
magnet reference track} has to satisfy all criteria for a reference
track in the pattern tracker~\cite{conc}, including the momentum
requirement $p>1$~GeV/c. In addition, the track is required to pass at
least four layers within the area of high magnetic field, defined by a
circle of 140~cm radius around the magnet center in the bending
plane. It is also demanded that the track pass at least one layer in
the first half of the magnet ($z < 450$~cm), to reject late decays
where the particles do not traverse a sufficient amount of integrated
field. Since it can happen that particles at the outer and inner
borders of the acceptance ``circumvent'' some superlayers, tracks with
such gaps are also regarded as not being covered by the geometrical
acceptance. Finally, tracks are not considered as belonging to the
reference set if they have lost more than 30\% of their energy within
the magnet. This requirement is directed against radiating electrons
which have a reduced probability to be selected by the first level
trigger.

\label{sec:proper}
A track candidate is regarded to properly reconstruct a particle if
at least 70\% of its hits were caused by this particle. In addition,
it is checked that the first reconstructed hit of the candidate is not 
further than 100~cm downstream of the first hit of the real
particle. The efficiency is then simply the fraction of reference
tracks which are successfully reconstructed. In order to isolate the
effect of the magnet propagation, we define in addition a {\it magnet
pattern recognition efficiency} which is defined {\it relative} to the 
track finding efficiency in the pattern tracker:
\begin{displaymath}
\epsilon_{magnet} = \frac{\epsilon_{magnet+pattern\ tracker}}
{\epsilon_{pattern\ tracker}}
\end{displaymath}
The efficiency $\epsilon_{pattern\ tracker}$ has been re-evaluated for
the geometry and simulated data sample used here and was found to
agree, within statistical errors, with the numbers presented
in~\cite{conc}.


\subsection{Geometrical acceptance}
\begin{figure}
        \epsfig{file=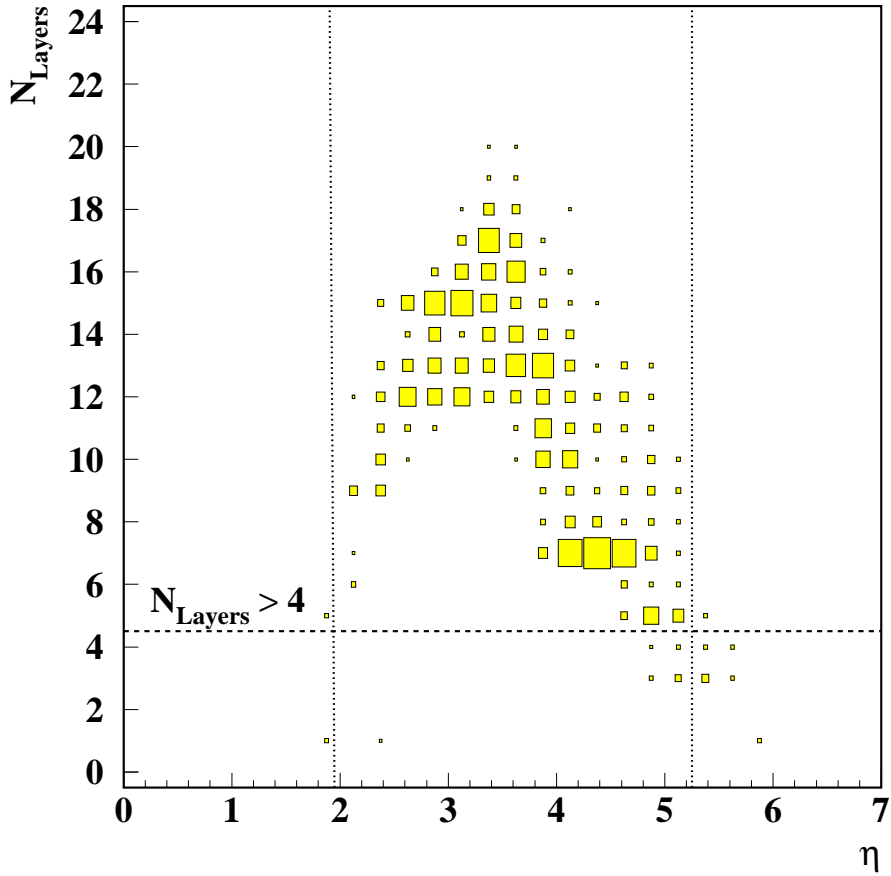,width=13cm}
        \caption{Number of layers passed in the magnet as a function
of pseudorapidity $\eta$. The different levels of layer count around
$\eta =3$ and $\eta =4.5$ reflect the different number of superlayers
of the outer and inner tracking systems. The vertical lines correspond 
to the nominal acceptance region as described in the text.}
        \label{accmag}
\end{figure}

In order to verify the adequateness of the reference track criteria,
fig.~\ref{accmag} shows the number of layers passed in the magnet, as
defined in the last section, as a function of pseudorapidity $\eta$
for muons from the golden B decay which are reference tracks in the
pattern tracker. In the outer tracker dominated regime around $\eta =
3$, the layer count is higher than in the region governed by the inner
tracker around $\eta = 4.5$. Obviously the cut $N_{Layers}>4$ gives a
generous description of the geometrical acceptance. The nominal
rapidity coverage~\cite{herabdesign}, corresponding to 10~mrad at the
inner and 250~mrad (horizontal) resp. 160~mrad (vertical) at the outer
border, is displayed as a pair of vertical lines in fig.~\ref{accmag}.

The fraction of particles which are selected as reference tracks can
be interpreted as the geometrical acceptance of the combined system of
pattern and magnet tracker. Table~\ref{tab:accmag} summarizes these
values for particles related to the golden B decay. Comparison of the
muon acceptance of 80.2\% with the one of the pattern tracker
alone~\cite{conc} shows that 92\% of the muons within the pattern
tracker acceptance have also enough hits in the magnet tracker. The
electron acceptance is smaller (66.6\%) because of the cut on the
radiated energy. The pion and $K^0_S$ acceptances are mainly limited
by the $K^0_S$ decay length which causes a considerable fraction of
kaons to decay too late for a good momentum measurement of the
pions. The total fraction of golden $B^0$ decays with reconstructable
condition is 34\% in the muon and 29\% in the electron channel.

\subsection{Efficiency}
\begin{figure}
\begin{center}
\unitlength1cm
\begin{picture}(13,19)
\put(0.0,9.5){\epsfig{file=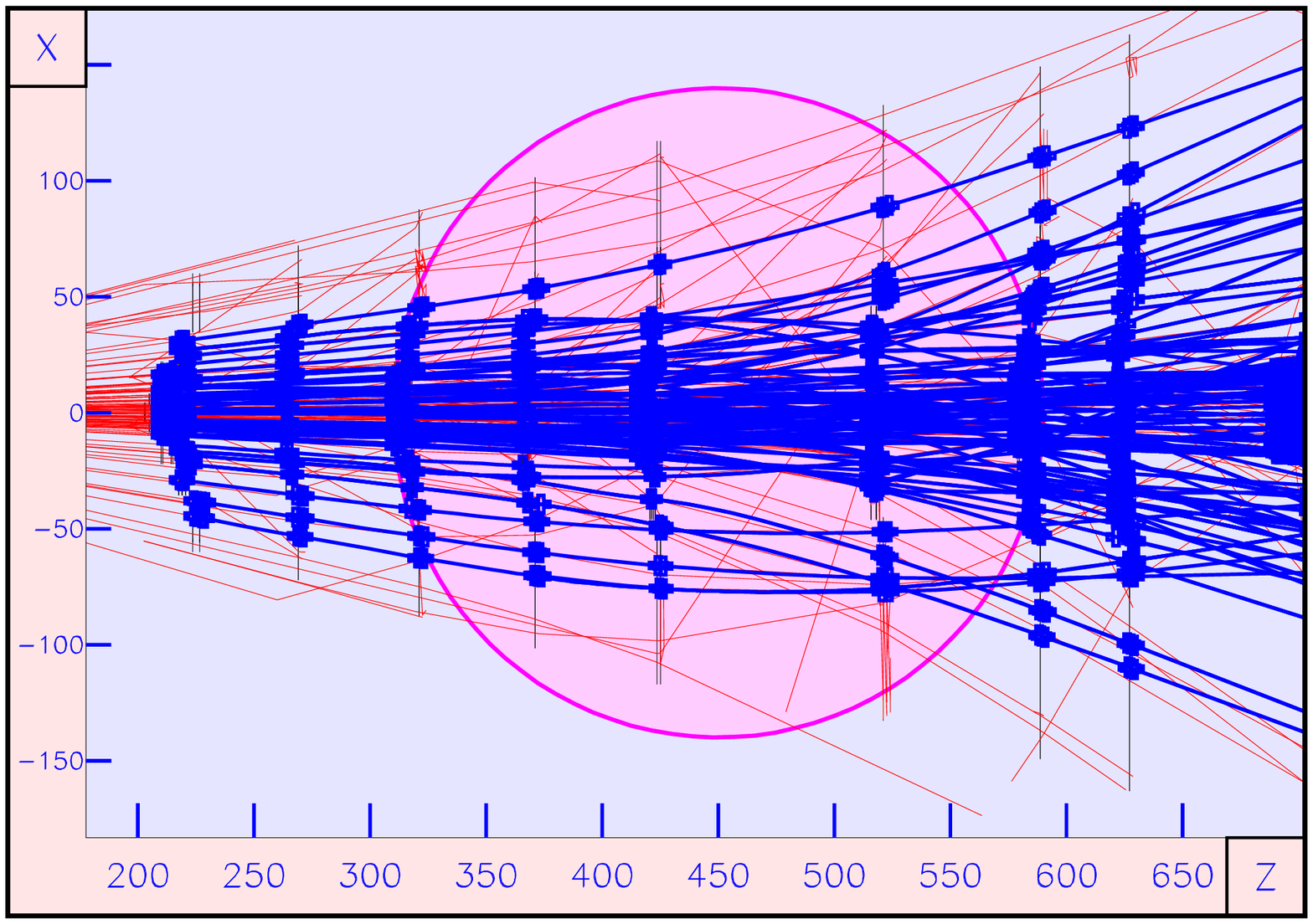,width=13cm}}
\put(1.9,18.1){\makebox(0,0)[t]{\huge (a)}}
\put(0.0,0.0){\epsfig{file=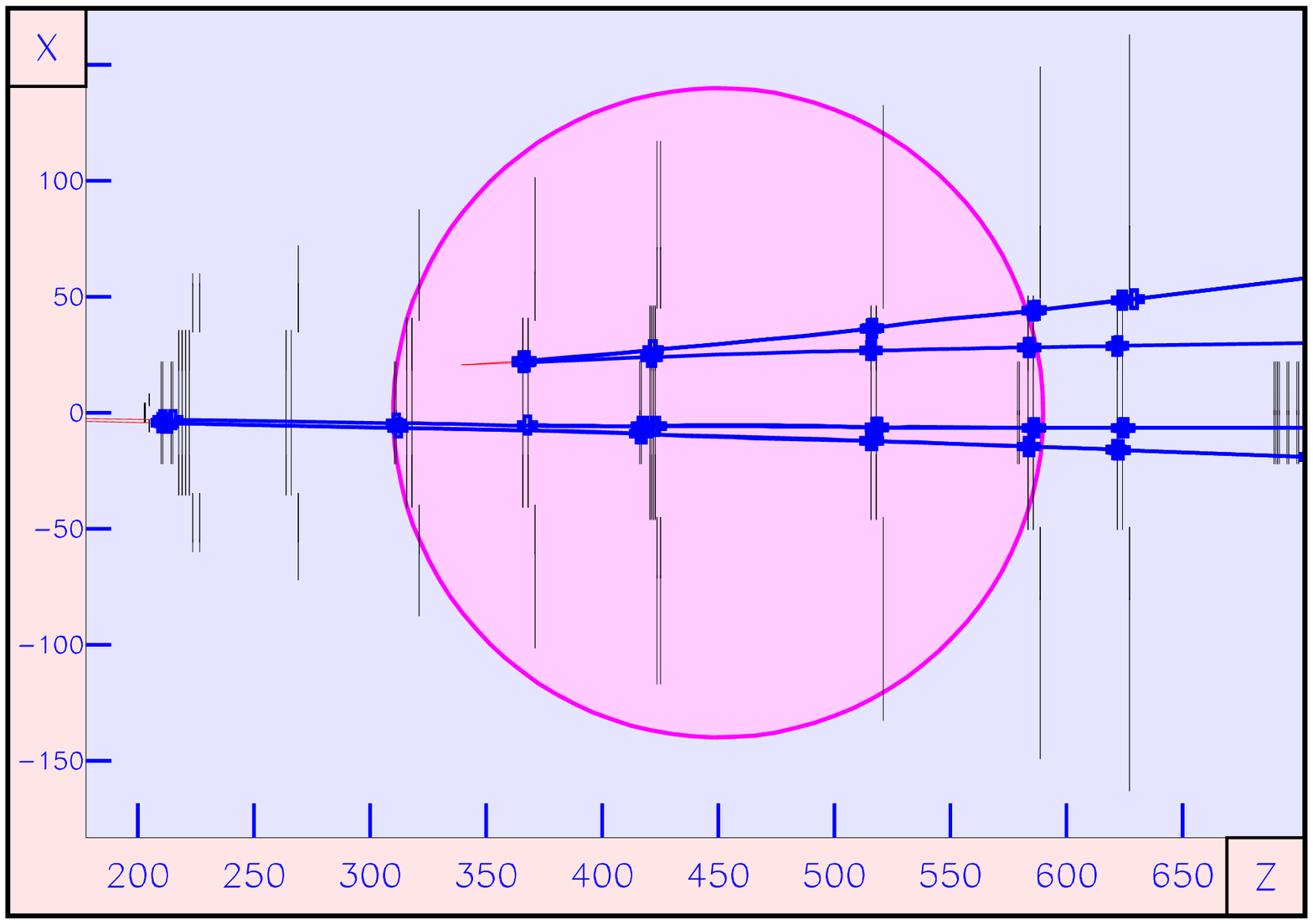,width=13cm}}
\put(1.9, 8.5){\makebox(0,0)[t]{\huge (b)}}

\put(12.3, 6.9){\makebox(0,0)[t]{\large $\pi^+$}}
\put(12.3, 6.2){\makebox(0,0)[t]{\large $\pi^-$}}
\put(12.3, 5.5){\makebox(0,0)[t]{\large $\mu^+$}}
\put(12.3, 4.6){\makebox(0,0)[t]{\large $\mu^-$}}
\end{picture}
\end{center}
\caption{(a) Display of an event with one interaction containing the
golden B decay and six superimposed inelastic interactions, focussed
on the magnet area. Both the Monte Carlo tracks (light grey) and the
reconstructed tracks (thick dark lines) are show (reconstructed hit
points denoted by crosses).  (b) Same event, with the display
restricted to particles from the golden $B$ decay.}
\label{mag-display}
\end{figure}

Figure~\ref{mag-display} shows displays of an event with seven
superimposed interactions with focus on the magnet tracker region.
The Monte Carlo tracks have been drawn by connecting the impact points
in the sensitive detector parts, note that because of this technique,
low-momentum particles have a poor representation. The region of
highest magnetic field which is used for the definiton of reference
tracks is shown as a filled circle.  While several softer particles
are not reconstructed because they do not reach the pattern tracker,
the vast majority of tracks passing the full spectrometer is found in
full length up to the entrance of the magnet tracker. The second
display (fig.~\ref{mag-display}b) is restricted to the particles from
the golden $B$ decay and shows the proper recognition of these tracks.

The pattern recognition efficiencies obtained in B interactions with
on average four superimposed inelastic interactions are summarized in
tab.~\ref{tab:effmag}.  The right column gives the relative efficiency
of the magnet propagation step. For charged particles from the golden
B decay, these efficiencies are larger than 97\%. The $\pi^\pm$
appears to have a slightly lower efficiency than the leptons which may
be attributed to the late decays of some $K^0_S$'s and the resulting
smaller number of hits, and the on average smaller momentum. Charged
particles above 1~GeV/c have a magnet propagation efficiency of
$\approx 94\%$.

The left column shows the efficiency of the combined pattern and
magnet tracker reconstruction. The ratio to the numbers in the right
column corresponds within statistical errors to the numbers given
in~\cite{conc}. It is interesting to note that, after the requirement
that reference tracks should not radiate more than 30\% of their
energy within the magnet, electrons and muons have similar
efficiencies. The relative rate of ghosts\footnote{relative to the
number of reference tracks} is 2\%, corresponding to
an average of 0.85 ghost tracks per event. This is much smaller than
the 4.4 ghosts per event which remained after pattern tracker
analysis~\cite{conc}, which shows the ghost rejection power of the
magnet tracker. Further suppression of ghosts is expected from
the matching with the vertex detector system.

If one includes the geometrical acceptance, the total fraction of
reconstructed golden B decays is 28\% in the muon and 24\% in the
electron channel.

\begin{figure}
\epsfig{file=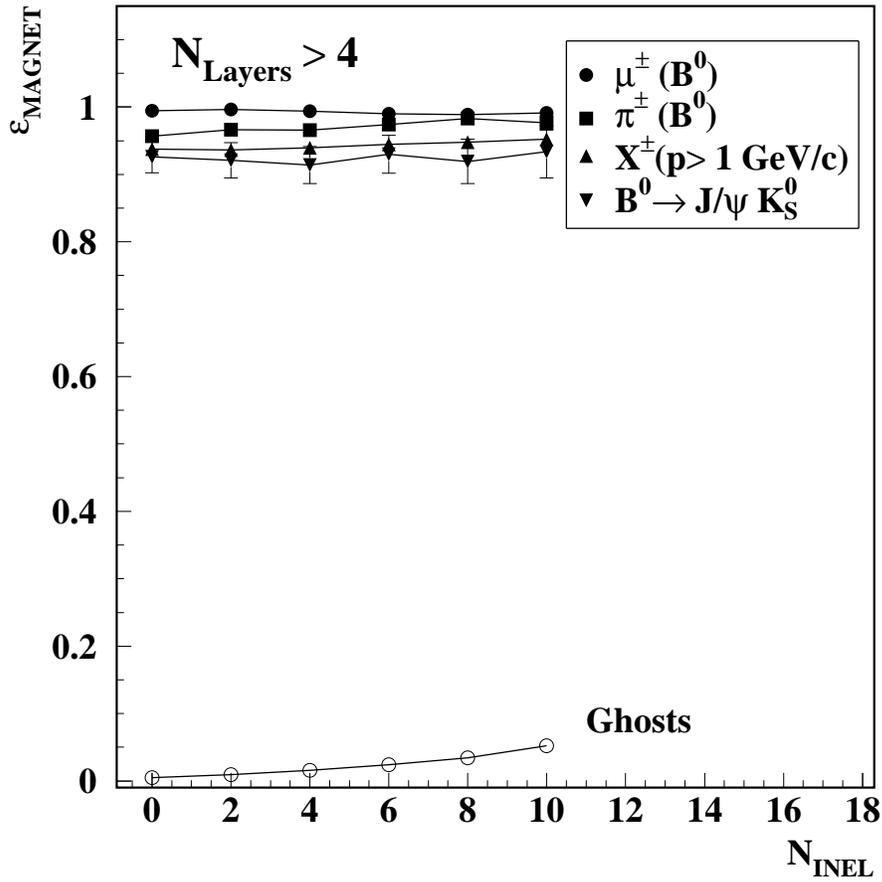,width=13cm} 
\caption{Pattern
recognition efficiency in the magnet versus number of
superimposed inelastic interactions, for particles which traverse
at least five layers in the central part of the magnet, and
have been detected as segments in the pattern tracker. The 
ghost rate is also shown.} 
\label{effmag-vs-nint}
\end{figure}

In order to probe the robustness of the algorithm, the efficiency of
the magnet tracker propagation step has been studied as a function of
the number of superimposed inelastic
interactions. Figure~\ref{effmag-vs-nint} shows that the magnet
tracking efficiency remains at the same high level up to ten
superimposed interactions, so that the total efficiency is governed by
the behaviour from the pattern tracker. The ghost rate increases in a
well-controlled way.

\begin{figure}
        \epsfig{file=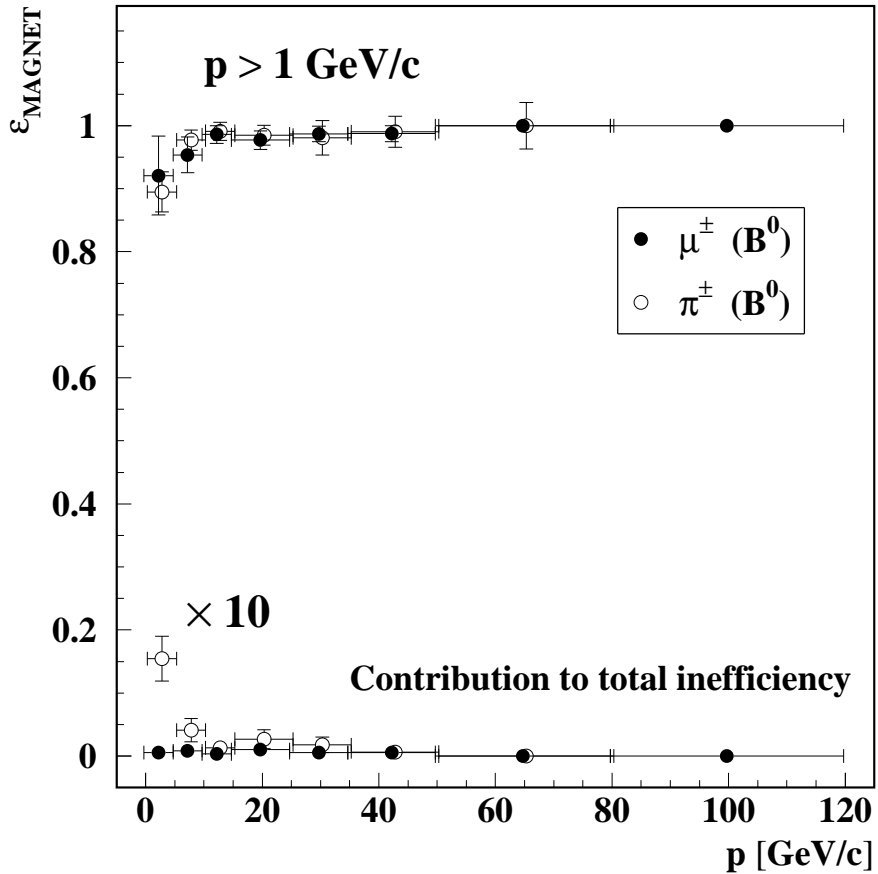,width=13cm}
        \caption{Magnet pattern recognition efficiency as a function
        of momentum for muons (filled circles) and pions (open
        circles) from the golden B decay. At the bottom, the
        contribution of each momentum bin to the mean 
        inefficiency $1-\epsilon_{Magnet}$ (multiplied by 10) is
        shown, indicating that the smaller pion efficiency
        is a result of the softer momentum spectrum.
}
        \label{effmag-vs-p}
\end{figure}

In order to investigate further the causes for particle loss, we have
broken down the magnet efficiency as a function of momentum for muons
and pions from the golden B decay (fig.~\ref{effmag-vs-p}). For both
particle species, the efficiency saturates on a high level above
5~GeV/c, below this value the efficiency drops, which is likely to be
due to multiple scattering effects, or to strongly curved trajectories
which are not entirely covered by the navigation tables. At the bottom
of fig.~\ref{effmag-vs-p}, it is indicated how the different parts of
the momentum range contribute to the mean inefficiency, $1 -
\epsilon_{Magnet}$. While muons 
and pions show the same efficiency at a given momentum, the pions
suffer more strongly from the reduced efficiency at lower momentum
because of their softer spectrum.

\begin{figure}
        \epsfig{file=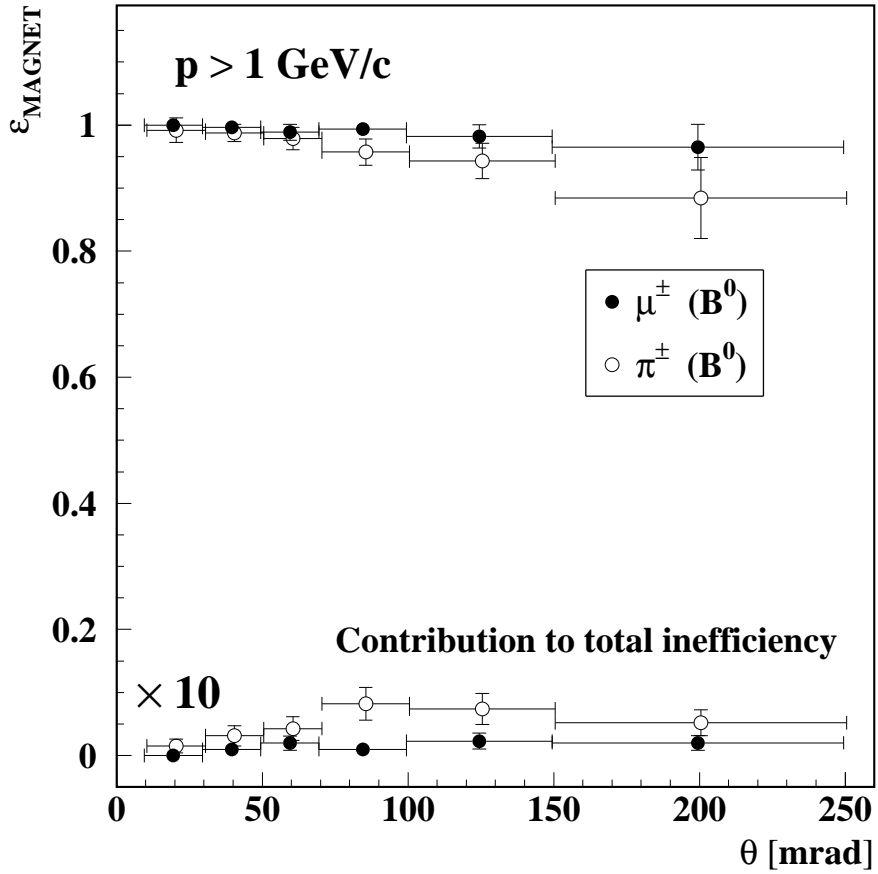,width=13cm} 
        \caption{Magnet pattern
        recognition efficiency as a function of polar angle for muons
        (filled circles) and pions (open circles) from the golden B
        decay. At the bottom, the contribution of each angle bin to
        the mean inefficiency $1-\epsilon_{Magnet}$ (multiplied by 10)
        is shown.}  
        \label{effmag-vs-angle}
\end{figure}

Figure~\ref{effmag-vs-angle} shows a similar distribution as a
function of the polar angle, measured at the track point of lowest $z$
within the magnet tracker. The magnet track finding efficiency
decreases visibly with increasing angle. Since, however, polar angle
and total momentum are strongly correlated within the HERA-B
kinematics, the effect is not independent from the one in
fig.~\ref{effmag-vs-p}.

\subsection{Parameter estimates}
The underlying reconstruction concept~\cite{conc} foresees that the
track candidates emerging from the magnet track reconstruction should
be matched with track candidates found in the vertex detector, or
continued into the latter by direct propagation, which is only
possible if the track parameters are well estimated. In principle, a
fully iterative refit of the trajectory to the hits found by pattern
recognition, which takes also additional traversed material into
account, can be used to optimize the parameter estimate. Such a fit,
the method of which is described in~\cite{rangerfit}, has been used in
the following for comparison. From the aspect of cpu time consumption,
it would however be preferred if a global refit can be postponed until
the hits from all components have been collected.

\begin{figure}
\begin{center}
\epsfig{file=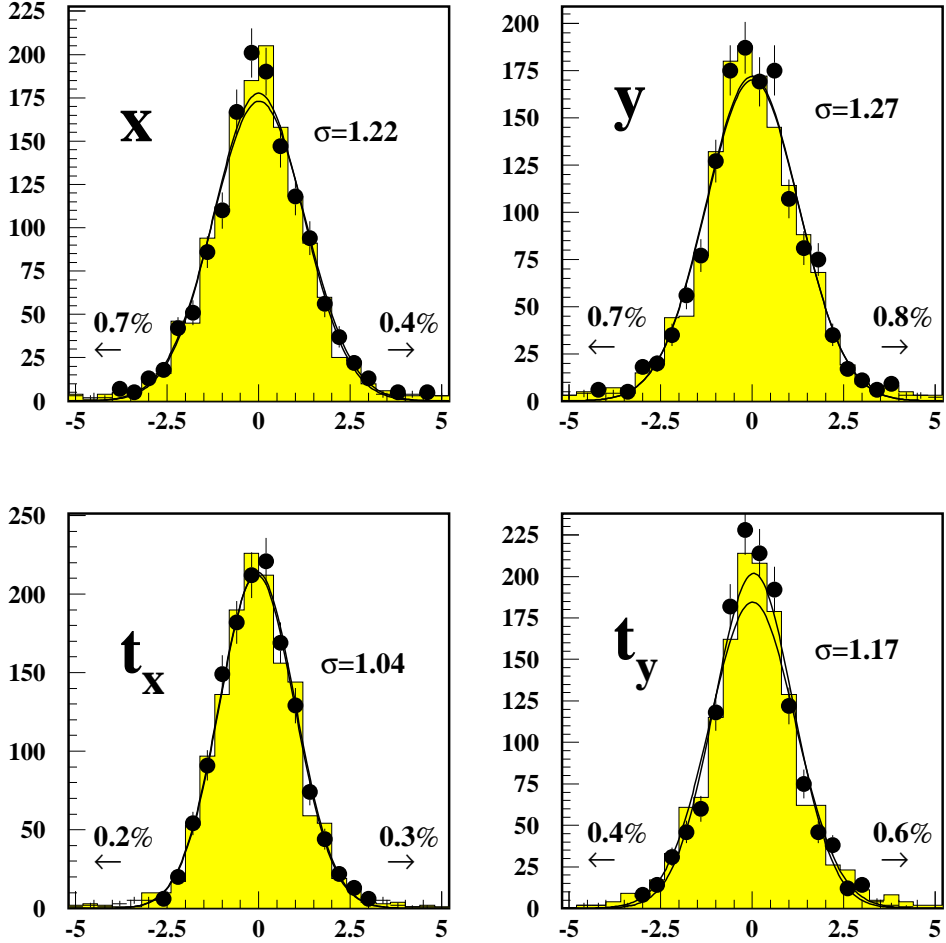,width=15cm}
\end{center}
\caption{Normalized residuals (see text for explanation) 
for the parameters $x$, $y$, $t_x$ and $t_y$ for muons from the golden
B decay. The solid points with error bars are directly obtained from
the parameters delivered by the magnet propagation, the shaded
histogram is the result of an iterative refit which interpolates
traversed material between the hits. The result of Gaussian fits to
both distributions is also shown, and the standard deviation $\sigma$
is quoted for the refitted case. The percentages at the arrows in the
lower parts of the figures give the number of entries below and above
five estimated standard deviations for the parameters delivered by the
refit.}
\label{pullxmag}
\end{figure}

Figure~\ref{pullxmag} shows the normalized residuals 
\begin{displaymath}
  \frac{{\cal X}_i^{REC} - {\cal X}_i^{MC}}
       {{\cal C}_{ii}^{1/2}}
\end{displaymath}
where ${\cal X}_i^{REC}$ is the reconstructed parameter at the point
of lowest $z$ of a track candidate, ${\cal X}_i^{MC}$ the
corresponding Monte Carlo-truth and ${\cal C}_{ii}$ the estimate for
the corresponding covariance matrix diagonal element, where the
parameters are ${\cal X}_i=$ $x$, $y$, $t_x$ and $t_y$. (If several
track candidates exist for the same particle, the better one is chosen
because it is more likely to be selected in the subsequent matching
with the vertex detector.) The error estimate ${\cal C}_{ii}^{1/2}$
which is used to normalize the residual is determined by the
coordinate resolution and the multiple scattering effects as they are
estimated during magnet propagation, or through the full
refit. Dilutions due to wrong hits or left/right assignments in the
pattern recognition process are expected to lead to widths larger than
one. All observed residual distributions have shapes close to
Gaussians. There are hardly any tails, as is also reflected in the
underflow and overflow percentages which are given in the diagrams,
which shows that the criterion used to label a track as {\it
reconstructed} (see section~\ref{sec:proper}) is adequate. The fitted
widths exceed the ideal value by 17--27\%, which can be attributed to
inevitable pattern recognition effects. Only in the case of the $t_x$
parameter the excess is smaller (4\%), which is not unexpected because
the resolution in this parameter is dominated by multiple
scattering. One concludes that the track parameters delivered by the
magnet propagation are good enough to be directly used by the matching
step without a refit at this stage.

\begin{figure}
\begin{center}
\unitlength0.9cm
\begin{picture}(12,19)
\put(0.0,18.5){\makebox(0,0)[t]{\huge (a)}}
\put(1.0, 9.5){\epsfig{file=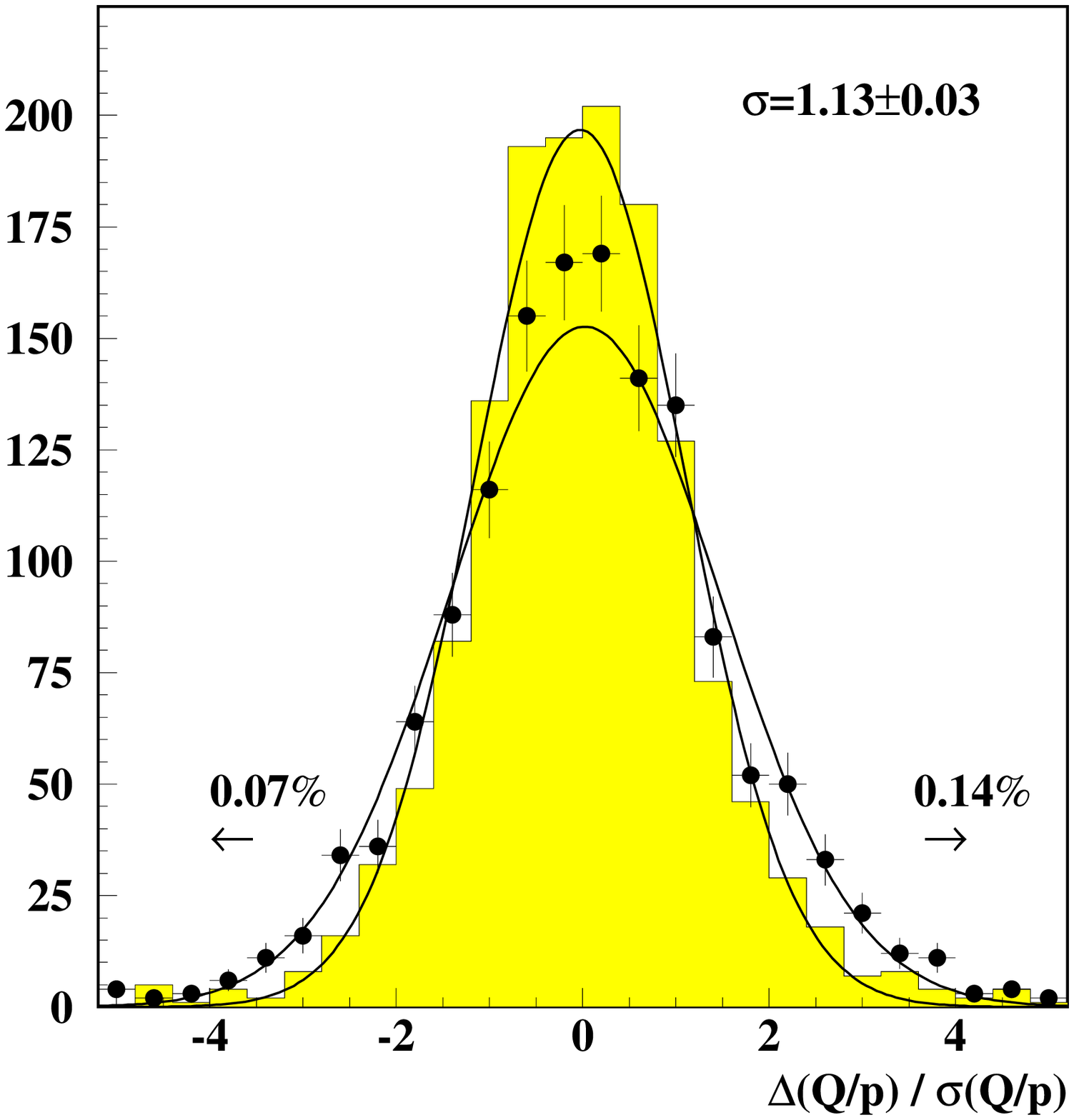,width=9cm}}
\put(0.0, 9.0){\makebox(0,0)[t]{\huge (b)}}
\put(1.0, 0.0){\epsfig{file=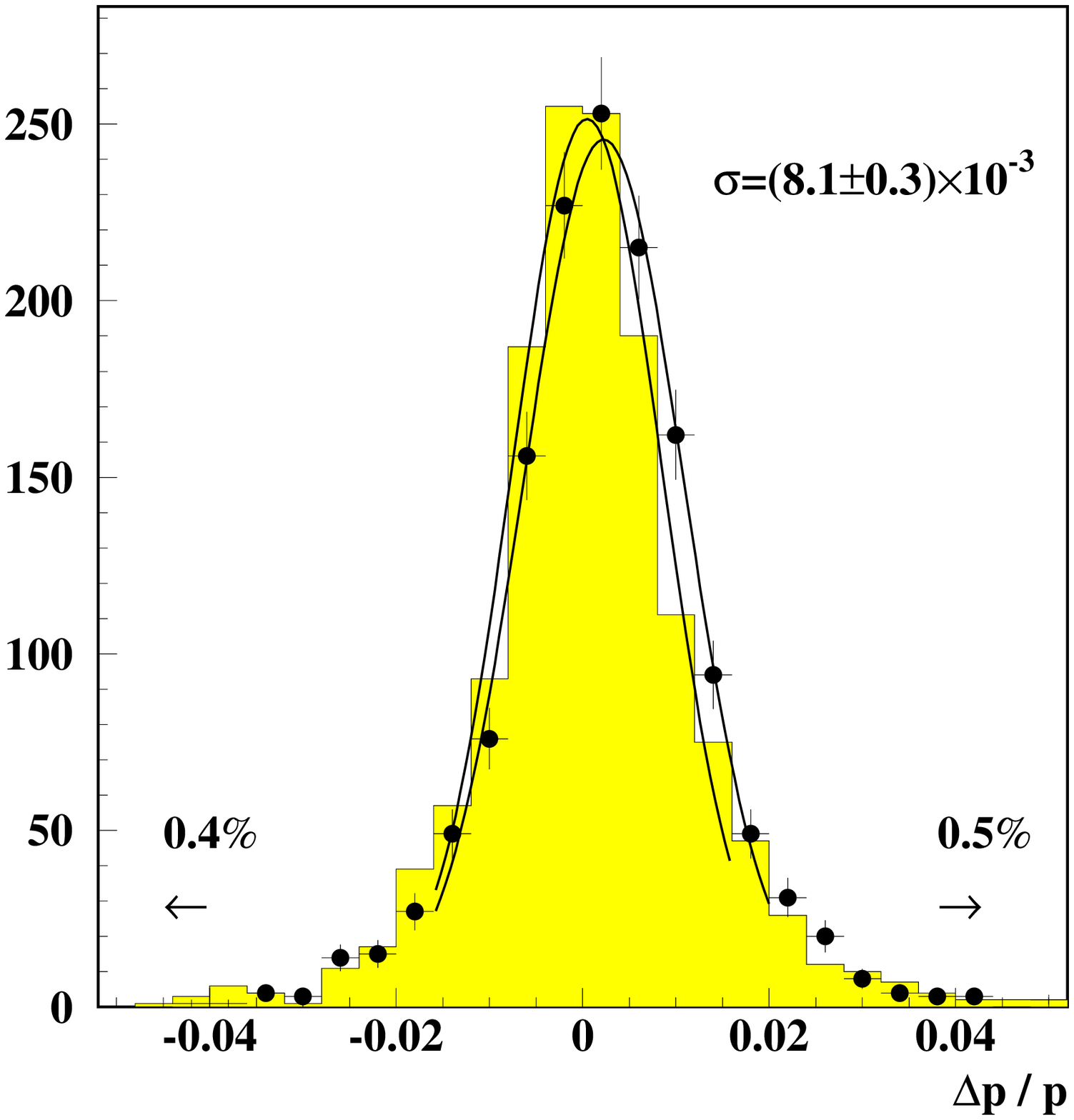,width=9cm}}
\end{picture}
\end{center}
\caption{Distributions of (a) the normalized residual for the
parameter Q/p, and (b) the relative momentum residual $\Delta p/p$ for
muons from the golden $B$ decay.  The solid points with error bars
correspond to the parameter as it is delivered by the magnet
propagation, the shaded histogram is the result of the iterative
refit. The results of Gaussian fits to both distributions (for (b) to
the central part) is superimposed, where the standard deviation
$\sigma$ is quoted for the refitted case.  }
\label{pullpmag}
\end{figure}

The quality of the momentum parameter estimate influences the matching
to the vertex detector segment only if the first point (at lowest $z$)
is well within the magnetic field, because otherwise a linear
extrapolation should be sufficient. Figure~\ref{pullpmag} shows in its
upper part the normalized residual of the momentum parameter ($Q/p$),
which does also show only moderate tails and a width only slightly
larger than unity.  The bottom part of the same figure displays the
distribution of the relative momentum error, $\Delta p / p$.  Note
that the distribution is not expected to be Gaussian because of the
momentum spread and multiple scattering effects.  The mean relative
resolution obtained by fitting the central part of this distribution
is $8.7 \cdot 10^{-3}$, which the refit improves slightly to $8.1
\cdot 10^{-3}$. To compare these resolutions with the technical limit,
the same events were passed through a full iterative refit of the hits
selected using Monte Carlo information\footnote{This procedure is
sometimes called {\it ideal pattern recognition}}, where a mean
resolution of $7.1 \cdot 10^{-3}$ was obtained. The resolutions are
summarized in table~\ref{tab:resol}.  The comparison thus gives an
illustration of the size of dilutions caused by pattern recognition
effects, which turn out to be relatively moderate. Further improvement
can be expected from
\begin{enumerate}
\item a rejection of outliers, based on the $\chi^2$ contribution of
each hit to the final fit, which can diminish the dilution from wrong
hits or left/right assignments.
\item successful matching with vertex detector segments, which should 
reduce the relative influence of wrong hit information within the
magnet.
\end{enumerate}
We conclude that the method studied here gives reasonable parameter
estimates in spite of the high track densities.

\subsection{Speed}
\begin{figure}
        \epsfig{file=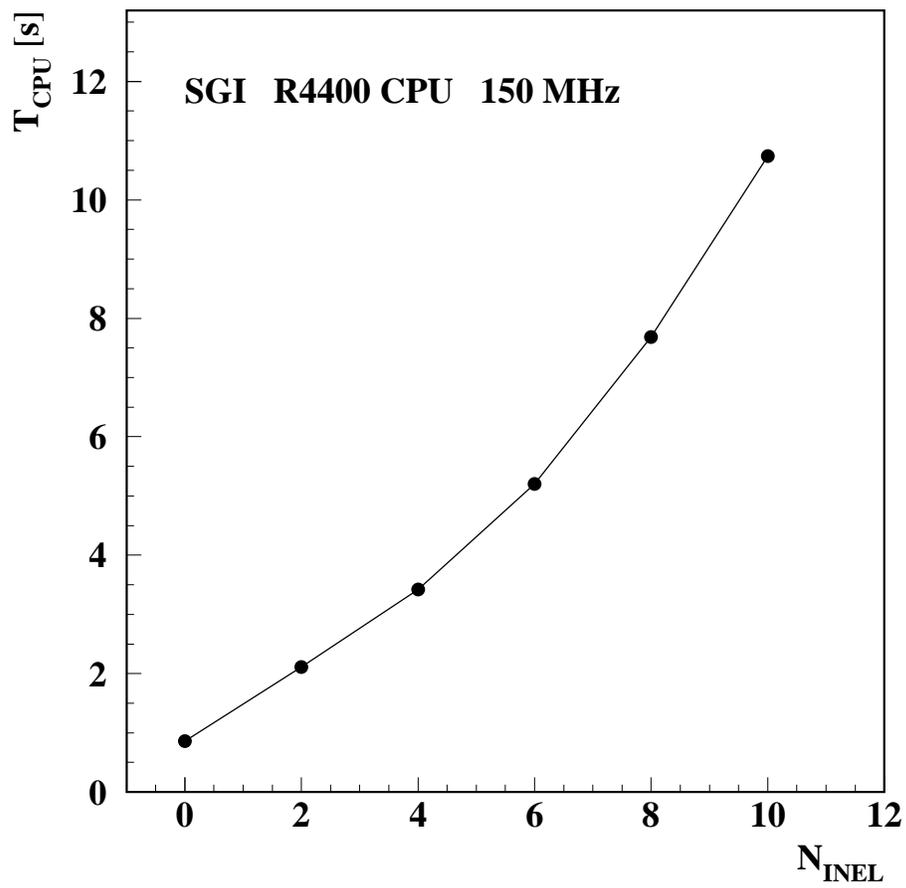,width=13cm}
        \caption{Computing time of the magnet propagation step per
        event as a function of the number of inelastic 
        interactions which are superimposed on the interaction
        generating the golden B decay.} 
        \label{cpumag}
\end{figure}

In the magnet case, the issue of computing time is potentially even
more critical than in the field-free case because of the costly
numerical parameter transport within the magnetic field. The time
consumption of the magnet part has been tested by running the
reconstruction algorithm with and without magnet propagation step and
computing the difference. The typical precision of such measurements
is $\approx 20\%$.  The resulting behaviour as a function of the
number of superimposed interactions in displayed in
fig.~\ref{cpumag}. On an SGI challenge workstation, the computing time
in case of on average four superimposed inelastic interactions is
about 3.6~s\footnote{It is expected that the processors of the HERA-B
reconstruction farm will be faster by at least a factor of 2.}. The
increase with the number of interactions is not much steeper than
linear, corresponding to an almost constant time requirement per
track. Again, the speed behaviour of the Concurrent Track Evolution
algorithm turns out to be favourable for online processor farm
applications, which was already observed in the field-free
system~\cite{conc}.

\section{Summary}
The Concurrent Track Evolution algorithm has been extended and applied
to the problem of associating detector hits with tracks in the
inhomogeneous magnetic field of a forward B spectrometer. The method
has been tested on simulated events with the currently implemented
HERA-B geometry. The efficiency for propagating tracks from the golden
B decay was found to be 97\% and better. The ghost rate from the
pattern tracker segments is considerably reduced by the magnet
propagation step. The resulting track parameters are good enough for
the vertex detector match without additional iteration or refit. The
cpu time consumption is smaller than for the initial pattern tracker
reconstruction step, and shows an uncritical behaviour with increasing
number of superimposed interactions.

\section*{Acknowledgments}
We would like to thank Thomas Lohse for contributing his tracking
expertise in many inspiring discussions, and Siegmund Nowak for his
continuous support regarding the HERA-B detector simulation
software. One of us (A.S.)  would like to thank DESY Zeuthen for the
kind hospitality extended to him during his visit.  This work was
supported by the Bundesministerium f\"ur Bildung, Wissenschaft,
Forschung und Technologie under contract number 05~7~BU~35I~(5).

\newpage

\section*{Tables}
\begin{table}[hbp]
  \begin{center}
  \begin{tabular}{|l|c|}
  \hline
  Particle        &  Geometrical Acceptance \\
  \hline
  \hline
  $\mu^{\pm}_{J/\psi}$     &  $(80.2 \pm 1.3)\%$  \\
  $e^{\pm}_{J/\psi}$       &  $(66.6 \pm 1.9)\%$  \\
  $\pi^{\pm}_{K^0_S}$      &  $(66.8 \pm 1.1)\%$  \\
  \hline
  $J/\psi \rightarrow \mu^+\mu^-$     &  $(63.4 \pm 1.6)\%$  \\
  $J/\psi \rightarrow e^+e^-$         &  $(45.4 \pm 2.9)\%$  \\
  $K^0_S \rightarrow \pi^+\pi^-$      &  $(54.8 \pm 1.6)\%$  \\
  \hline
  $B^0 \rightarrow J/\psi K^0_S \rightarrow \mu^+\mu^-  \pi^+\pi^-$ 
                                      &  $(34.0 \pm 1.6)\%$  \\

  $B^0 \rightarrow J/\psi K^0_S \rightarrow e^+e^-  \pi^+\pi^-$
                                      &  $(28.6 \pm 2.6)\%$  \\
  \hline
  \end{tabular}
  \end{center}
  \caption{Geometrical acceptance for particles related to the
                golden  B decay.}
  \label{tab:accmag}
\end{table}

\begin{table}[hbp]
\centering
\begin{tabular}{|l||c|c|}
\hline
Particle & $\epsilon$ & $\epsilon$ \\
         & Pattern + Magnet & Magnet \\ 
\hline
\hline
$e^\pm_{J/\psi}$   & (96.8 $\pm$ 0.9)\% & (98.7 $\pm$ 1.1)\% \\
$\mu^\pm_{J/\psi}$ & (97.3 $\pm$ 0.4)\% & (99.6 $\pm$ 0.6)\% \\
$\pi^\pm_{K^0_S}$  & (93.3 $\pm$ 0.7)\% & (97.4 $\pm$ 0.9)\% \\
\hline
$J/\psi \rightarrow e^+e^-$         &  $(93.5 \pm 2.1)\%$  
                                    &  $(97.3 \pm 2.6)\%$  \\
$J/\psi \rightarrow \mu^+\mu^-$     &  $(94.5 \pm 0.9)\%$  
                                    &  $(99.3 \pm 1.3)\%$  \\
$K^0_S \rightarrow \pi^+\pi^-$      &  $(88.0 \pm 1.5)\%$  
                                    &  $(94.8 \pm 1.9)\%$  \\
\hline
$B^0 \rightarrow J/\psi K^0_S \rightarrow e^+e^-  \pi^+\pi^-$
                                    &  $(83.9 \pm 3.9)\%$   
                                    &  $(92.3 \pm 4.9)\%$  \\
$B^0 \rightarrow J/\psi K^0_S \rightarrow \mu^+\mu^-  \pi^+\pi^-$ 
                                    &  $(83.0 \pm 2.1)\%$   
                                    &  $(93.6 \pm 2.9)\%$  \\
\hline
${\rm X^\pm (p>1~GeV/c) }$ & (88.1 $\pm$ 0.2)\% & (94.1 $\pm$ 0.2)\% \\
{\rm Ghosts}       & (2.0$\pm$ 0.1)\% &                    \\
\hline
\end{tabular}
\caption{Pattern recognition efficiency for ordinary particles
and those related to the golden B decay. The relative ghost rate is 
also shown.}
\label{tab:effmag}
\end{table}

\begin{table}[hbp]
\centering
\begin{tabular}{|l||c|c|}
\hline
Method       & \multicolumn{2}{|c|}{$\Delta p/p$} \\
             & Gaussian fit  $\sigma$    & FWHM \\
\hline
magnet propagation  & (8.66 $\pm$ 0.27) $\cdot 10^{-3}$ 
& (20.4 $\pm$ 0.6) $\cdot 10^{-3}$ \\
\hline
magnet propagation  &               &             \\
~~~~~~+ refit     & (8.08 $\pm$ 0.29) $\cdot 10^{-3}$ 
& (19.0 $\pm$ 0.7) $\cdot 10^{-3}$ \\
\hline
ideal pattern recognition &          &              \\
~~~~~~+ fit       & (7.12 $\pm$ 0.22) $\cdot 10^{-3}$ 
& (16.8 $\pm$ 0.5) $\cdot 10^{-3}$ \\
\hline
\end{tabular}
\caption{Relative momentum resolution $\Delta p/p$ determined both
from fitting a Gaussian to the central peak of the residual
distribution (middle column), and as the full width at half maximum
(right column). The resolution is given as (a) directly obtained from
the magnet propagation procedure, (b) from a refit to the hits
returned by this procedure and (c) from a fit to the hits selected
using Monte Carlo information ({\it ideal pattern recognition +
fit}).}
\label{tab:resol}
\end{table}

\clearpage  

\newpage

\section*{Figures}

\end{document}